# Too much of a good thing? Entrepreneurial orientation and the non-linear governance effects of SaaS platforms

**Authors**: Jacopo Ballerini [a,b,*], Magali Pino [a,c], Michal Kuděj [d], Alberto Ferraris [a,e]

[a] University of Turin, Department of Management, Corso Unione Sovietica 218 bis, 10134 Torino, Italy

[b] Vrije Universiteit Amsterdam, Department of Marketing, De Boelelaan 1105, 1081 HV Amsterdam, the Netherlands

[c] IAE Lyon, Magellan, Université Lyon 3, Lyon, France

[d] Prague University of Economics and Business, Faculty of Business Administration, Department of Strategy, W. Churchill Sq. 1938/4, 130 67 Prague, Czech Republic

[e] Gnosis: Mediterranean Institute for Management Science, School of Business, University of Nicosia, 2417, Nicosia, Cyprus

## Abstract

Purpose:

This study examines the role of entrepreneurial orientation (EO) in the governance of software as a service (SaaS) platforms and its impact on strategic alignment and long-term governance performance in small and medium-sized enterprises (SMEs). By conceptualising SaaS as a hybrid governance model, this research investigates how transaction cost attributes affect strategic alignment and how EO moderates these relationships.

Design/methodology/approach:

The research employs a multi-study research design. Using partial least squares structural equation modelling (PLS-SEM) with reflective constructs' measurement, Study 1 analysed survey data from 180 UK and US entrepreneurs. Study 2 complemented these results with a quasi-experimental design that relied on a secondary dataset from 238 European start-ups, refining the operationalisation of the variables through industry-based indicators.

Findings:

The results reveal an inverted U-shaped relationship between human asset specificity and the frequency of SaaS use, and their alignment with SMEs' strategic objectives. Risk-taking enhances the alignment between human asset specificity and SaaS strategic alignment, while proactiveness strengthens the link between SaaS strategic alignment and long-term performance. Both studies support the idea that SaaS strategic alignment follows an inverted U-shaped relationship with long-term performance, indicating that excessive dependence on SaaS may negatively affect governance-enabled strategic outcomes over time.

Originality:

This study contributes to the entrepreneurial and digital governance literature by conceptualising SaaS as a hybrid governance model and investigating how SaaS impacts SMEs strategically, drawing on the perspectives of transaction cost theory and EO. It extends research by unveiling the non-linear effects of SaaS adoption on strategic alignment and performance, highlighting the role of entrepreneurial decision-making in digital technology adoption.

## 1. Introduction

The adoption of certain digital technologies underpinning the digital transformation context has driven competitive advantages across industries (Lerro *et al.*, 2022; Schiuma *et al.*, 2022), sustaining the long-term strategic performance of small and medium-sized enterprises (SMEs) (Costa *et al.*, 2024; Escoz Barragan and Becker, 2024). However, it has also created considerable implications for firms (Vial, 2019; Wu and Pambudi, 2024), impacting the types of investments and commitments they make (Beynon *et al.*, 2021; Cuypers *et al.*, 2021). Some of these technologies have been notably transformative (Braune *et al.*, 2025), and this is the case for the most popular cloud delivery model: Software as a Service (SaaS). SaaS solutions – particularly those adopted by UK and US firms – reached $145 billion in 2022 and their adoption is steadily increasing (Business Standard, 2020; Gartner, 2021).

SaaS is replacing the traditional 'owned, hosted and managed' model of business application software, also known as the 'on-premises' model. Compared to the the latter, SaaS is a cloud computing model owned, hosted and managed by external providers and distributed to the firm through the Internet as a service on a recurring fee basis (Benlian and Hess, 2011; Loukis *et al.*, 2019; Marston *et al.*, 2011; Senyo *et al.*, 2018). This emerging model enables SMEs to access secure, high-quality IT services, such as those found in larger organisations, at an affordable cost and scale, which may support SMEs' core strategic outcomes (Attaran and Woods, 2019). However, despite compelling evidence of value propositions for firms in adopting the SaaS model (Dutta *et al.*, 2024), specific peculiarities – such as the shift of particular tasks and responsibilities to external providers and short-term contracts to utilise standardised services – generate distinctive governance challenges which lead to the need to re-examine SaaS adoption and performance (Schneider and Sunyaev, 2016).

While mainstream research rooted in the transaction cost theory has historically conceptualised governance mechanisms in binary terms – namely, 'hierarchies' relying on vertically integrated control and 'markets' relying on outsourcing (Argyres and Zenger, 2012; Cuypers *et al.*, 2021; Williamson, 1975) – recent findings reveal that SaaS multiparty and multifaceted contexts give rise to hybrid governance mechanisms that blend control and coordination across firms' boundaries. These 'hybrid' mechanisms are characterised by complex managerial and entrepreneurial implications, including blurred boundaries of authority and accountability (Attaran and Woods, 2019; Cuypers *et al.*, 2021; Lerro *et al.*, 2022; Park *et al.*, 2017).

However, existing studies on SaaS adoption tend to investigate structural and environmental drivers and inhibitors, such as technical readiness and regulatory pressures, without fully addressing complex interpersonal dynamics and strategic orientations at the firm level (Oliveira *et al.*, 2019; Yang *et al.*, 2015). Even advanced models, such as the risk-based IT governance framework proposed by Dutta et al. (2024), which focus on both risk perception associated with SaaS adoption and strategic decision-making alignment depending on the firm's appetite for risk, do not incorporate firm-level strategic postures, such as entrepreneurial orientation (EO). Yet, previous research assumed that greater EO, such as risk-taking and proactivity, can support SMEs in achieving higher performances in a digitalised context (Hervé *et al.*, 2020; Kraus *et al.*, 2023; Niemand *et al.*, 2021). Therefore, not considering EO in relation to SaaS adoption (Loukis *et al.*, 2019) identified as a 'hybrid' form of governance (Cuypers *et al.*, 2021) is critical, as it overlooks how firms with comparable governance arrangements may adopt SaaS in divergent ways due to differences in their EO strategic posture, which may lead to diverse performances (Morris *et al.*, 2011). Investing in this issue is of interest to scholars and practitioners, as the results should offer them guidance on the role of entrepreneurial and business factors in determining the performance consequences of the adoption and management of SaaS solutions (Attaran and Woods, 2019).

In this study, performance is not conceptualised purely from the financial or short-term perspectives. Rather, we focus on long-term governance performance, defined as the effectiveness of SaaS-based IT governance mechanisms in supporting SMEs' strategic objectives over time. This perspective is consistent with the transaction cost theory lens adopted and with prior research on IT governance and strategic alignment. Therefore, this paper aims to answer the following naturally arising research questions (RQs): Considering the adoption of SaaS as a 'hybrid' form of governance, what is its impact on SMEs' performances? (RQ1), and how does an SME's entrepreneurial orientation (EO) affect the adoption and performance of the SaaS model? (RQ2). To address these RQs, we develop hypotheses grounded in the perspective of transaction cost theory (TCT) (Cuypers *et al.*, 2021) and EO (Covin and Wales, 2012). The researchers empirically test the hypotheses using a mixed-method approach. Study 1 consists of a survey-based inquiry distributed to 228 entrepreneurs living in the UK and the USA who adopt SaaS platforms as part of their business operations. Study 2 employs a quasi-experimental research design, utilising a dataset of 238 European start-ups. The findings show that, first, the relationship between two key attributes, such as human asset specificity and frequency, and the alignment of the adopted SaaS model with the firm's core business strategy is explained by a quadratic inverted relationship (inversed U-shaped) and in the case of human asset specificity it is moderated by the risk-taking dimension of the EO construct, while in the case of frequency it is (partially) moderated by the proactivity dimension of EO. Second, the relationship between SMEs' SaaS model strategic alignment and the firm's long-term performance is also explained by an inverted U-shaped relationship moderated by the EO's proactivity dimension.

This study brings novelty and development to the digital transformation and entrepreneurship literature by pioneering the investigation of the relationship between SaaS platforms and SMEs' core business performance as a hybrid mode of governance. The paper also considers the 'hybrid' form of governance proposed by transaction cost theory, positioning EO as a moderator, thus introducing a novel contingency perspective: the effectiveness of governance-enabled strategic alignment in promoting SaaS adoption is shaped by the firm's EO.

## 2. Literature review

### 2.1 SaaS as a hybrid governance arrangement: implications for strategic outcomes

In the early 2010s, the SaaS model emerged as a new way to supply software services, surpassing the application service provisioning (ASP) model (Benlian and Hess, 2011; Rossignoli et al., 2017). This new model can improve SMEs' governance-related strategic outcomes (Attaran and Woods, 2019). Indeed, the accessibility of SaaS solutions via the Internet eliminates the need to install and run applications on client firms' computers (Marston et al., 2011; Senyo et al., 2018; Venkatachalam et al., 2014), which makes them more reliable, efficient, and available 24/7 (Venkatachalam et al., 2014; Waters, 2005). Also, SaaS multi-tenant architecture (i.e. all clients share a standard source code and computing resources) (Senyo et al., 2018; Song et al., 2020) reduces the total cost of ownership and mitigates risks in terms of technology upgrades insulation (Benlian and Hess, 2011; Waters, 2005).

However, despite the potential benefits of SaaS adoption by SMEs, it also raises IT governance challenges (Aubert and Rivard, 2020; Benlian and Hess, 2011; Venkatachalam et al., 2014). Throughout the paper, the term 'governance' refers specifically to IT governance in the context of SaaS adoption, understood as the allocation of decision rights, accountability, and control mechanisms related to the use and management of SaaS platforms (Weill and Ross, 2004). It does not refer to corporate governance in a broader ownership or board-related sense. Theoretical foundations for understanding these governance dynamics stem from the transaction cost theory (TCT), which initially proposed two discrete alternative governance modes: hierarchies and markets (Williamson, 1975). The former rely on administrative control, while the latter rely on price mechanisms. In the case of SaaS solutions, an external provider manages the the virtual infrastructure and future developments (Benlian and Hess, 2011; Senyo *et al.*, 2018), while the using, setting, and customising is done by the firm itself (Schneider and Sunyaev, 2016). SaaS solutions establish a context in which external entities make a growing number of IT strategy decisions (Aubert and Rivard, 2020; Park *et al.*, 2017) leading to a model somewhere between outsourcing and complete vertical integration (Loukis, Janssen, and Mintchev 2019): a 'hybrid form of governance'. Such governance form remains underinvestigated (Cuypers et al. 2021; De Vita, Tekaya, and Wang 2010), and we lack a clear enough understanding of it to predict SMEs' governance-related outcomes (Cuypers *et al.*, 2021; Loukis *et al.*, 2019) despite the vital importance of such knowledge for SMEs (Braune *et al.*, 2025).

## 2.2 Entrepreneurial orientation as a contingency in SaaS governance effectiveness

In addition to governance issues, the SaaS model presents significant management and structural challenges that affect the value that SMEs can derive from SaaS solutions (Alemayehu *et al.*, 2023; Benlian and Hess, 2011; Venkatachalam *et al.*, 2014). Specifically, Costa *et al.* (2024) unveils three influential dimensions: the culture, resources, and management of SMEs. Alemayehu *et al.* (2023) further clarify that SaaS implementation quality alone is insufficient to generate performance benefits; over 70% of digital technology projects fail to meet their objectives resulting in the wastage of billions of euros (Tabrizi *et al.*, 2019). This can be explained by the insufficient consideration of firm specificities (e.g. industry-related context) and maturity (Braune *et al.*, 2025; Costa *et al.*, 2024; Wu and Pambudi, 2024), or the misalignment of management practices with the firm's digital capabilities (Alemayehu *et al.*, 2023; Escoz Barragan and Becker, 2024).

To avoid these failures, firms need to adjust their strategy and management practices to create value through digital alignment by enhancing their digital orientation (Escoz Barragan and Becker, 2024; Kindermann *et al.*, 2021; Tabrizi *et al.*, 2019). Digital orientation refers to the firms' approach to pursuing digital technology-enabled opportunities to gain competitive advantages (Kindermann *et al.*, 2021). It combines market, learning, and entrepreneurial orientation (EO) (Quinton *et al.*, 2018).

EO is a firm's strategic posture rather than what it does (Anderson *et al.*, 2015; Covin and Lumpkin, 2011; Lumpkin and Dess, 2001) and can be considered through three dimensions: risk-taking; proactiveness; and innovativeness (Kraus *et al.*, 2023; Putniņš and Sauka, 2020; Wales *et al.*, 2021). Risk-taking is the willingness to invest lots of resources with a reasonably high chance of failure (Miller, 1983); proactiveness is an opportunity-seeking position that involves anticipating and seizing future market demands ahead of the competition (Lumpkin and Dess, 2001); while innovativeness is symbolised by firms' strong commitment to take technological leadership (Covin and Slevin, 1991). By combining proactive and innovative entrepreneurial behaviours with a managerial attitude, firms can seize opportunities with uncertain outcomes (Anderson *et al.*, 2015).

Recent findings further reinforce this view by showing that EO may not only be a driver of digital success but also a strategic response to navigate increasingly complex environments (Escoz Barragan and Becker, 2024) and face governance and structural challenges that digital technologies (like SaaS) carry with them; thus supporting SMEs in achieving stronger governance-enabled outcomes (Niemand *et al.*, 2021).

In light of this, we argue that prior studies on EO paired with the TCT perspective can extend the literature's theoretical bases (Park *et al.*, 2017) to investigate the potential moderating role of EO in SaaS adoption. Specifically, EO may influence how firms navigate the constraints and opportunities presented by the SaaS model as a hybrid form of governance, and in turn, shape the impact of SaaS adoption on SMEs' core business performance.

### 3. Theoretical background

Built on the original work of Coase (1937) and Williamson (1985), the transaction cost theory (TCT) has become one of the most influential theories with regard to management and information systems (Ballerini *et al.*, 2024; Cuypers *et al.*, 2021). Originally used to make vertical integration decisions (i.e. "make versus buy"), TCT has been applied to explore diverse organisational boundary decisions like horizontal diversification, strategic alliances, or supply chain relationships. TCT recommends that firms choose their governance structure – 'markets', 'hierarchies', or 'hybrids'– based on the differences in transaction costs (Cuypers *et al.*, 2021; Geyskens *et al.*, 2006). Three dimensions – asset specificity, uncertainty, and transaction frequency – are used to observe transactions and predict governance structure efficiency (Williamson, 1975). 'Asset specificity' refers to the degree to which different parties can reuse an asset (e.g. human capital, skills, or brand name capital) in alternate transactions without sacrificing productive value (Williamson, 1985). The more specific the asset is, the higher the associated expenses for redeploying it will be (Tadelis and Williamson, 2013). Uncertainty can be defined as an unpredicted disturbance in a transaction for which unforeseen adaptations are needed. The effect on generic transactions is minimal but is of importance when applied in non-vertical integrated context (Cuypers *et al.*, 2021; Tadelis and Williamson, 2013; Williamson, 1985). As for the 'transaction frequency' (i.e. transaction recurrence) (Williamson, 1975, 1985), it has ambiguous implications. On the one hand, one can argue that if a transaction has a low recurrence, implementing a specialised internal structure may not be cost-effective. Alternatively, if it recurs frequently, it is possible to recover the costs of developing a specialised management infrastructure. On the other hand, frequently recurring contracts suggest that future business opportunities are at risk, – making a good reputation an important factor, and thus becomes a part of the comparative contractual considerations (Tadelis and Williamson, 2013).

Knowing that, 'markets' is the less costly governing structure for generic asset transactions with negligible hazards while the 'hierarchical' approach (i.e. vertical integration) should be considered when transactions become frequent, uncertain, and asset-specific (Cuypers *et al.*, 2021; De Vita *et al.*, 2010; Tadelis and Williamson, 2013). For 'hybrid' contracting arrangements at the crossroads between generic market transactions and hierarchy, the two polar models relying on price mechanism (i.e. 'markets') and administrative control (i.e. 'hierarchies') have fewer advantages. Thus, a 'hybrid' governance structure can be defined as 'an intermediate mode of organisation that uses credible commitments to support exchange for transactions that pose an intermediate range of hazards' (Tadelis and Williamson 2013, p. 36).

Leveraging this seminal literature on TCT, we can identify the adoption of SaaS models as the employment of 'hybrid' corporate governance in the functional domain where the SaaS platform is used. We reiterate that SaaS models differ from complete outsourcing and vertical integration in that an external provider develops the application and manages its hosting, structure, technical side, and updates, while the firm uses, sets, and customises it (Schneider and Sunyaev, 2016). Therefore, we believe that TCT can address the gaps identified in the literature review.

### 4. Hypotheses development

#### 4.1 Transaction costs theory attributes and SaaS adoption strategy and performance

In line with TCT rationale, insourcing is more cost-efficient and creates advantageous strategic benefits if the provision of firms' application services requires a high level of human asset specificity (Dibbern *et al.*, 2005). Indeed, following the TCT rationale, the more employees involved in a specific project accumulate knowledge (i.e. high human asset specificity) the more that resource becomes indispensable to the firm, encouraging vertical integration (Cuypers *et al.*, 2021; Tadelis and Williamson, 2013; Williamson, 1985). On the contrary, human knowledge is to be considered a low-specificity asset as long as it can be redeployed across alternative projects and by alternative users without a significant loss in productive value. In such a scenario, horizontal integration is to be preferred (Williamson 1996).

In the case of digital technologies in general, and SaaS in particular, the literature reveals that firms must cultivate and develop human knowledge over time to achieve distinctive performance from these technologies (Gupta *et al.*, 2024; Hervé *et al.*, 2020). Thus, to be consistent with TCT reasoning (Argyres and Zenger, 2012), one can argue that the more firms invest in specialised human resources to manage a strategic digital tool, the more appropriate it becomes to retain direct control over that tool. However, since SaaS solutions are not fully controlled by client firms (Benlian and Hess, 2011), this 'hybrid' governance model may be suboptimal for managing highly core functionalities that require high asset specificity. Instead, SaaS may be better suited for supporting less strategic functions with less human capital investment. Our reasoning leads us to formulate the following hypothesis:

> H1: There is a quadratic inverted relationship (inverted U-shaped) between human asset specificity (HAS) and SaaS alignment with SME strategy

Another significant factor in predicting and explaining SaaS alignment with SMEs' core strategy is 'frequency'. Indeed, the extent to which users adopt and utilise a technology is crucial to understanding its implementation outcomes (Aggarwal and Kryscynski, 2015; De Vita *et al.*, 2010). Nowadays, computing is omnipresent and SaaS promises to provide software packages for a wide range of applications (Ferrari *et al.*, 2012; Marston *et al.*, 2011; Zhang *et al.*, 2016). While some software is dedicated to day-to-day usage (i.e. not specific to the firm's core business), such as accounting, billing, or emails (Ma and Seidmann, 2015), others are for core-business functions such as PayPal or Shopify (Dushnitsky and Stroube, 2021; Marston *et al.*, 2011). These core-business function-specific SaaS are based on per-transaction fees; the more frequent the transactions, the more the client firms have to pay for them (Ma and Seidmann, 2015).

Moreover, compared to an ad hoc solution, the SaaS model gives client firms less control over the solution evolution (Benlian and Hess, 2011; Senyo *et al.*, 2018). In a sense, it echoes the relational governance mechanisms where clients and vendors mutually define and adjust routines without an explicit pattern of authority over future developments (Colombelli *et al.*, 2019). This may have two consequences: 1) an increase in coordination complexity between the client firm and the SaaS solution provider as the usage increases (Dibbern, 2003); and 2) a shift between the SaaS solution provider strategy and the client firm's core strategy (Benlian and Hess, 2011).

Therefore, based on TCT rationale, we argue that SaaS adoption (as a hybrid form of governance), could be suitable for firms approaching a new activity with increasing attention up to a certain threshold. Once the activity becomes a highly frequent strategic task that drives the firm's value proposition and competitive advantages, the need to develop in-house solutions to reduce costs and simplify the governance arises. Therefore, we drafted the following hypothesis:

> H2: There is a quadratic inverted relationship (inverted U-shaped) between frequency and SaaS alignment with SME strategy

TCT states that uncertainty represents a key determinant of governance choice, as it captures the extent to which transactions are affected by unpredictable changes that require unforeseen adaptations (Williamson, 1985). However, in the specific context of digital technologies and SaaS solutions, uncertainty primarily manifests through technological uncertainty rather than market or managerial factors (Kessler et al., 2022; Song & Montoya-Weiss, 2001). The rationale of TCT (Williamson, 1975, 1985) infers that hybrid relational structures should be favoured in the absence of technological uncertainty (De Vita et al., 2010). In such situations, hybrid models like SaaS (Loukis et al., 2019) should allow firms to capture some of the vertical integration benefits (i.e. lower transaction costs) while also benefiting from the economic advantages of market transactions, including cost savings and value creation through outsourcing (De Vita et al., 2010). However, SMEs adopting SaaS face challenges stemming from the vendor's technological road map opacity, the pace of software updates, the external the dependency on external infrastructures, and the lack of transparency regarding data migration or interoperability (Benlian and Hess, 2011; Schneider and Sunyaev, 2016). These conditions expose firms to unforeseen adaptation costs that directly affect the SaaS model's alignment with the firm's strategic objectives.

Focusing on technological uncertainty is thus consistent with both TCT and the nature of the SaaS model, which hinges on technological dependencies and evolving architectures rather than market turbulence. Consequently, we argue that as technological uncertainty increases, SMEs are less likely to rely on SaaS solutions for core strategic activities, preferring instead to limit their use to non-core or supporting functions. This reasoning leads us to the following hypothesis:

> H3: There is a negative relationship between technological uncertainty and SaaS alignment with SME strategy

Finally, firms are not relying solely on SaaS to improve internal efficiency, but also to target more strategic business capabilities and drive business innovation through digital affordances (Berman *et al.*, 2012). Digital affordances originate from digital infrastructures' technical architecture. They support the redesign of value creation, delivery, and capturing (Autio *et al.*, 2018). In the case of SaaS, this redesign allows: business scalability; cost flexibility (firms pay for what they use); market adaptability (regarding processes, services and products); context-driven variability (expanded computing capacity); ecosystem connectivity; and end-user masked complexity (Benlian and Hess, 2011; Berman *et al.*, 2012; Hervé *et al.*, 2020; Waters, 2005). However, behind this masked complexity, SaaS 'hybrid' governance prompts organisational challenges to harness its innovative power and co-evolutionary potential (Berman *et al.*, 2012; Schneider and Sunyaev, 2016) and constrains client firms' main functionality customisation options (Benlian and Hess, 2011). This hinders SaaS advantages for firms (Benlian, 2009) and may lead to a risk of misalignment with client firms' core strategy over time (Benlian and Hess, 2011), coupled with platform dependency due to solution switching costs (Cutolo and Kenney, 2021).

Therefore, we believe that, to a certain degree, SaaS can align with firms' core strategy, sustain their competitive advantage, and thus ensure their long-term governance performance. However, the SaaS client firms' desire for new versions may not align over time (Benlian and Hess, 2011; Gupta *et al.*, 2024). Thus, it is reasonable to expect that at the beginning, when selecting an SaaS platform, SMEs will choose one that is most aligned with their core strategy to support their competitive advantages. However, over the succession of versions, the SaaS platform may diverge from the SME's core strategy (which may evolve itself). Thus, if SMEs' competitive advantages depend too strongly on SaaS, their long-term performance may decline over time. Therefore, we formulate the following hypothesis:

H4: There is a quadratic inverted relationship (inverted U-shaped) between SaaS alignment with SME strategy and long-term governance performance

### 4.2 Entrepreneurial orientation moderation effects

Entrepreneurial orientation (EO) is of interest in investigating the impact of SaaS on client firms' core business performance (Escoz Barragan and Becker, 2024). Indeed, as seen previously, research has generally established a positive relationship between EO aggregated measures and firm performance. Specifically, risk-taking has been frequently investigated (Kreiser *et al.*, 2013; Putniņš and Sauka, 2020; Wiklund and Shepherd, 2011). In the case of digital technologies, we consider both financial and operational risks, as it may be difficult to align them with traditional routines and misalignment may affect firms' performances (Kessler *et al.*, 2022; Yang and Yee, 2022). A deeper understanding of these risks may help mitigate them and enhance the quality of entrepreneurial decision-making. To do so, training (Tipu, 2017) and experiments seem to be interesting tools (Beyer *et al.*, 2016). Thus, in the case of SaaS adoption, we argue that the more the firm is willing to take risks in SaaS investment, train its employees, and conduct SaaS implementation experiments, the more it will be able to achieve a good alignment between SaaS and SME's core strategy. This assumption leads to the following hypothesis:

> H5: Risk-taking has a positive moderating effect on the relationship between human asset specificity and SaaS alignment with SME strategy

Although risk-taking can bring important benefits to the firm, it may highly depend on the technological uncertainty surrounding the projects (Tipu, 2017; Wiklund and Shepherd, 2011). Indeed, in an uncertain situation, the probabilities of future outcomes are unknown (Knight, 1921; Putniņš and Sauka, 2020) and knowledge is key to making enlightened decisions (Hervé *et al.*, 2020). Against this background and the discussion in Hypothesis 3, we argue that in highly uncertain situations, the more SMEs take risks, the greater the chance they will act like gamblers. Specifically, we assume that the more uncertain the situation is, the higher the risk of SMEs making mistakes in aligning SaaS with their core business strategy. Based on this assumption, we formulated Hypothesis 6:

> H6: Risk-taking negatively moderates the relationship between technological uncertainty and SaaS alignment with SMEs' strategy

In an uncertain environment, SMEs' deliberate engagement (i.e. proactivity) in capitalizing on emerging market opportunities is likely to boost performances via its positive effect on the level of risk-taking (Kreiser *et al.*, 2013). Indeed, proactive behaviour generally supports SMEs in discovering and exploiting opportunities ahead of their competitors (Smith and Cao, 2007), giving them a strategic edge (Hervé *et al.*, 2020). We propose a similar logic for SaaS adoption. SaaS tools may have applications that tend to be more support-orientated than core business functionality-orientated (Ma and Seidmann, 2015), but a particularly proactive attitude among SME managers can lead them to test functionality or customise applications to the best of their ability, thereby enhancing the extent to which those applications can also contribute to the specific core business strategy of the firm. This effect may especially flourish in frequently used SaaS tools. Therefore, we formulate the following hypothesis:

> H7: Proactivity positively moderates the relationship between frequency and SaaS alignment with SMEs' Strategy

Finally, previous research has established a beneficial relationship between the implementation of digital solutions and a firm's ability to identify new opportunities (Kraus *et al.*, 2023; Maravić *et al.*, 2022; Niemand *et al.*, 2021). In particular, Kraus *et al.* (2023) have unveiled that: 1) there is a positive relationship between the deployment of a digitalisation strategy and disruptive innovation; and 2) EO (i.e. firms' proactiveness, innovativeness, and willingness to take risks) has a positive effect on the digital strategy deployment. In the case of cloud computing, such as SaaS, competitive advantages for SaaS clients derive from the change in how information is delivered and used (Ojala, 2016). However, the first step for SMEs is to be able to detect and leverage these opportunities.

Therefore, building on previous discussions on proactivity, as developed in Hypothesis 6, it is conceivable that the more SMEs are proactive in identifying opportunities revealed by the benefits of the implemented SaaS platform, the more they can uncover long-term opportunities for their business. This reasoning leads to formulating the following hypothesis:

> H8: Proactivity positively moderates the relationship between SaaS alignment with SME strategy and long-term governance performance

**FIG. 1: HYPOTHESISED FRAMEWORK**

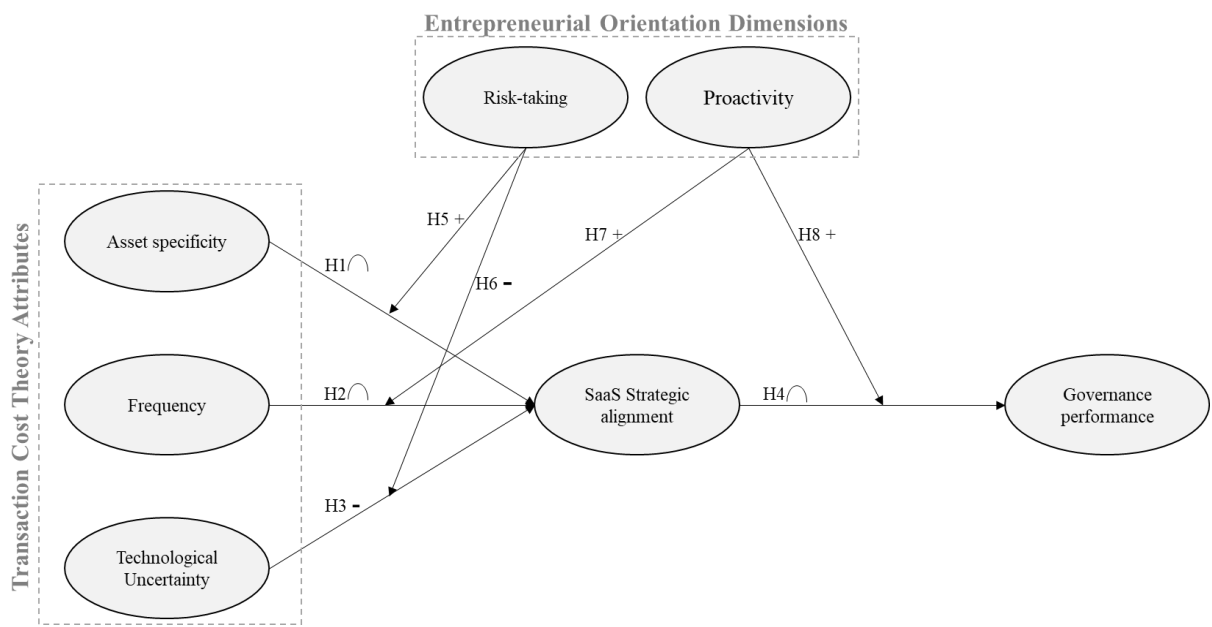

## 5. Research design

To address the research questions, we adopt a **multi-study quantitative research design** combining complementary analytical logics. Multi-study designs are particularly suited to theory-driven research where complex relationships benefit from being examined across different empirical settings and operationalisations, thereby strengthening robustness and inferential confidence. As noted in foundational methodological work, combining multiple studies and designs enables researchers to assess whether observed relationships hold beyond a single data structure or measurement approach (Brady and Collier, 2010; Jick, 1979).

In line with this perspective, Study 1 tests the hypothesised relationships using survey-based primary data and perceptual measures of governance-enabled performance grounded in IT governance theory. Study 2 complements this analysis by adopting a different design and an alternative, objective operationalisation of performance based on firm-level growth indicators. Taken together, the two studies enable us to assess the **robustness and generalisability** of the proposed non-linear relationships across complementary performance measurement logics, without relying on a single empirical or operational lens.

## 6. Study 1

Given the exploratory yet theoretically grounded nature of our research, Study 1 adopts a survey-based quantitative approach to test the hypothesised relationships within the proposed framework (Fig. 1). This design allows us to capture governance-enabled performance outcomes through perceptual measures that are particularly suited to assessing strategic alignment and coordination effects.

### 6.1 Measurements

Firstly, the authors identified different possible measurement approaches for the factors they wanted to measure, as suggested by the literature. Referring to the constructs related to the TCT, we found items from De Vita, Tekaya, and Wang (2010) suitable for measuring HAS, from Song and Montoya-Weiss (2001) for measuring technological uncertainty (TU) and from Crook et al. (2013) for measuring frequency (FR). Secondly, we identified the items to measure the two constructs related to EO, namely Proactiveness and Risk-Taking, from the measurement scales proposed by Covin and Wales (2012). Third, to measure the strategic alignment (SA) between the adopted SaaS and business core strategy, we relied on Wu et al.'s (2015) measurement approach. The measurement of this construct involved first assessing how much the firm's strategy consisted of being a cost leader, a product differentiator, or having a strong brand identity on a scale of 1 to 5. Respondents were then asked to rate (again on a scale of 1 to 5) how much the deployment of Saas platforms allows their firms to benefit in each of the three types of strategy. A high 'strategic alignment' score is calculated when the functionality of an SaaS platform aligns more closely with the business's core strategy. As an example, if a respondent rates 2 out of 5 that his/her business's core strategy is to differentiate itself from competitors, and likewise rates 2 out of 5 that its Saas platforms allow it to differentiate itself from competitors, the strategic alignment level, in this case, will be calculated as the highest in terms of the firm's differentiation strategy. Finally, to measure performance, we adopt and adapt the scale proposed by Ali and Green (2012), which captures long-term IT governance performance. This construct reflects respondents' perceptions of how effectively SaaS-related governance mechanisms support strategic objectives, coordination, and decision-making over time, rather than short-term financial outcomes. All constructs were specified as reflective, as variables such as entrepreneurial orientation, asset specificity, or strategic alignment represent latent behavioural dispositions that cannot be decomposed into measurable components but are instead reflected through respondents' perceptions (Covin and Wales, 2012; Hair et al., 2019). The complete items' questions are shown in the Appendix B. Third, we tested the psychometric quality of the scales, which are detailed in the following sections. Additional control variables potentially explaining some of the impact on strategic alignment and SaaS governance performance, such as firms' size and revenues, have been considered, decreasing the threat of unobserved heterogeneity issues. Both the control variables were coded on a five-point scale.

## 6.2 Data collection

The survey was initially designed in English and reviewed by a manager of an SaaS company to obtain external feedback from an industry expert. After minor adjustments, it was finalised and ready to be distributed. We administered the questionnaire on the survey platform Prolific. We took several precautions to safeguard the reliability of our data. First, in order to ensure the right sample population fit, the respondents were pre-screened using the platform's filters with the following criteria: i) they declared themselves entrepreneurs running their business venture; ii) they should have at least one subordinate; iii they should be resident in the USA or UK. These UK and US entrepreneurs were intentionally selected because they operate in the world's most mature and comparable SaaS markets, thereby reducing institutional heterogeneity and linguistic inconsistency. Second, to ensure that all respondents represented an optimal target firm, we included three different questions in the questionnaire that directly and indirectly evaluated whether they were adopting any SaaS platform to run their business. Only respondents who replied positively to all three questions were retrieved for the survey. In addition, to prevent possible common method biases typical of survey data collection, we took several procedural remedies recommended by Podsakoff et al. (2003) that we detail further. We complied with methodological and psychological separation principles by using different scales and techniques to measure different constructs, including reverse scales, 1–5 and 1–7 scales, and never mentioning the ultimate purpose of the research. Furthermore, we explicitly ensured the anonymity of the respondents by adopting a professional platform, such as Prolific, to select the sample, and a professional tool, such as Qualtrics, to collect responses. This ensured that respondents' anonymity was guaranteed and clarified that we would not retain any of their personal data, thereby preventing possible social desirability biases. Additionally, we used a random order of administering the questions, avoiding the framework considered in the study, and relying on Qualtrics' randomisation functionality to primarily minimise the risk of convergent and discriminant validity issues. Finally, consistently adhering to the literature (Podsakoff *et al.*, 2003), we adopted the following criteria for clarifying the questions carried out for each item by: having (a) defined a priori ambiguous terms such as 'SaaS'; (b) avoided vague concepts and carried out examples, if necessary (e.g. explicitly mentioned examples of SaaS); (c) kept questions simple and concise; (d) avoided double-barrelled questions; (e) decomposed broad questions into multiple more specific ones; (f) avoided complex syntax. The detailed items per factor and measurement scales are reported in Appendix B. The survey was submitted to a total of 228 entrepreneurs. We eliminated one respondent who declared that he was not adopting any SaaS platforms and the other 47 respondents who did not pass all three attention checks included in the survey, resulting in a final sample of 180 respondents. The detailed characteristics of the respondents are presented in Table I.

**TABLE I: SAMPLE CHARACTERISTICS**

| Variable | Category | Frequency | Percentage |
| --- | --- | --- | --- |
| **Industry Sector** | Consultancy | 15 | 8.3% |
| | Education | 14 | 7.8% |
| | Entertainment | 12 | 6.7% |
| | Accounting & Finance | 15 | 8.3% |
| | Healthcare | 27 | 15.0% |
| | Information technology | 43 | 23.9% |
| | Manufacturing | 20 | 11.1% |
| | Other | 34 | 18.9% |
| **Revenue** | Less than 1M USD | 101 | 56.1% |
| | 1–5M USD | 41 | 22.8% |
| | 5–10M USD | 19 | 10.6% |
| | 10–25M USD | 11 | 6.1% |
| | 25+M USD | 8 | 4.4% |
| **Employees** | 1–5 | 63 | 35.0% |
| | 6–10 | 34 | 18.9% |
| | 11–25 | 24 | 13.3% |
| | 25–50 | 28 | 15.6% |
| | 51+ | 31 | 17.2% |
| **Country** | United Kingdom | 60 | 33.3% |
| | United States | 120 | 66.7% |
| **Gender** | Female | 70 | 38.9% |
| | Male | 110 | 61.1% |
| **Educational Level** | High school diploma/A-levels | 7 | 3.9% |
| | Technical/community college | 14 | 7.8% |
| | Undergraduate degree (BA/BSc/other) | 65 | 36.1% |
| | Graduate degree (MA/MSc/MPhil/other) | 78 | 43.3% |
| | Doctorate degree (PhD/other) | 16 | 8.9% |

## 6.3 Results

As a first step, we proceed with the reliability and validity assessment suggested in the literature (Hair *et al.*, 2019). In the case of reflective variables, the individual reliability of each item is ensured. For this purpose, the factor loadings on their latent variables are examined. All the loadings of our reflective constructs are above the 0.7 thresholds suggested in the literature (Hair *et al.*, 2019), with the only exception of PERF5 (.518), which we do not retain and SA3 (.612), which we retain in the model, ensuring an adequate average variance extracted (AVE) for the SA construct. Secondly, the reliability of the constructs is analysed using composite reliability indicators, which are generally supposed to be above 0.7 in exploratory research. In all cases, our model's indicators are higher than 0.75. In addition, convergent validity has been ensured by analysing the AVE. Consistently, all indicators exceed the minimum level of 0.5, indicating high levels of convergent validity for the constructs. Discriminant validity was assessed using the heterotrait–monotrait (HTMT) ratio. All HTMT values were below 0.90, except for three marginal cases, which are theoretically justified given the constructs' conceptual relatedness, especially when these exhibit low variance inflation factor (VIF) levels (< 3), thereby reducing the risk of multicollinearity. In line with Voorhees et al. (2016), such minor exceedances do not heavily threaten discriminant validity. Nonetheless, we conducted robustness checks with minor measurement refinements (Appendix A) that yielded consistent results. Table II summarises the items' loadings, AVEs, VIFs, and composite reliabilities. In contrast, the HTMT ratios are reported in Table A2 of Appendix A. Regarding the assessment of the structural model, it has been subjected to the standardised root mean square residual (SRMR) criterion. The SRMR value of .064 indicates a good fit, as it falls below the 0.8 threshold (Hair *et al.*, 2019).

**TABLE II: ITEMS SUMMARY**

| Latent Constructs | Retrieved from: | Items N | Factor Loadings | VIF | Ave | Composite Reliability |
|---|---|---|---|---|---|---|
| Human Asset Specificity (HAS) | (De Vita *et al.*, 2010) | HAS1 | .915 | 1.342 | .747 | .855 |
| | | HAS2 | .810 | 1.342 | | |
| Frequency (FR) | (Crook *et al.*, 2013) | FR1 | .902 | 1.509 | .790 | .883 |
| | | FR2 | .875 | 1.509 | | |
| Technological Uncertainty (TU) | (Song and Montoya-Weiss, 2001) | TU1 | .822 | 1.205 | .706 | .827 |
| | | TU2 | .858 | 1.205 | | |
| Proactivity (PROAC) | (Covin and Wales, 2012) | PROAC1 | .849 | 1.573 | .634 | .838 |
| | | PROAC2 | .790 | 1.401 | | |
| | | PROAC3 | .747 | 1.317 | | |
| Risk-Taking (RT) | (Covin and Wales, 2012) | RT1 | .815 | 1.548 | .646 | .845 |
| | | RT2 | .808 | 1.424 | | |
| | | RT3 | .788 | 1.370 | | |
| Strategic Alignment (SA) | (Wu *et al.*, 2015) | SA1 | .735 | 1.151 | .507 | .753 |
| | | SA2 | .778 | 1.126 | | |
| | | SA3 | .612 | 1.107 | | |
| Governance Performance (PERF) | (Ali and Green, 2012) | PERF1 | .830 | 2.290 | .690 | .918 |
| | | PERF2 | .809 | 2.339 | | |
| | | PERF3 | .826 | 1.983 | | |
| | | PERF4 | .862 | 2.531 | | |
| | | PERF5 | N.R. | N.R. | | |
| | | PERF6 | .824 | 2.017 | | |

*Note: N.R. stands for not retained*

The authors used bootstrapping with 5,000 subsamples and two-tailed 95% confidence intervals to evaluate the statistical significance of the model's path coefficients. Table II shows the results and significance of the effects analysed using the SMARTPLS4 software. The PLS-SEM results show that there exists a significant positive direct relationship between HAS and SA ($\beta = .224$; $p = .024$), but there also exists a stronger quadratic relationship with a more substantial statistical significance showing a lower p-value between HAS and SA ($\beta = -.149$; $p = .016$). Therefore, we could infer that the strongest relationship between HAS and SA is quadratic, supporting H1. The model supports the significance of a quadratic relationship between FR and SA ($\beta = -.173$; $p = .005$) even more neatly, since it does not show any direct relationship significance, supporting H2. The last hypothesis (H3), which relates the transaction cost theory construct, such as TU with SA, despite being negative, does not provide a significant result and is not supported. The other direct relationship of the investigation foresees the existence of a quadratic one between SA and long-term governance performance (PERF). Despite no significant linear relationship, the model reveals a significant quadratic relationship ($\beta = -.124$; $p = .002$), supporting H4.

To assess the robustness of the moderating effects and to mitigate potential multicollinearity concerns observed in the discriminant validity assessment, we complemented the PLS-SEM results with a mean-centred principal component analysis (PCA) approach. This additional test enables a reduction in shared variance among latent constructs, thereby increasing the granularity of the interaction terms. The PCA-based moderation analysis produced results that were directionally consistent with those of the original PLS-SEM model, with slight variations in significance levels. In both estimation approaches, risk-taking (RT) significantly moderates the relationship between HAS and SA, supporting H5 ($\beta = .215$, $p < .01$; $\beta = .097$, $p < .01$). Conversely, the moderating effect of RT on the TU and SA relationship (RT × TU → SA) remains non-significant in both models, failing to support H6. The moderating effect of proactiveness on frequency (PROAC × FR → SA) becomes slightly significant in the PCA model ($\beta = .084$, $T = 1.773$, $p < .07$), providing partial support for H7. Finally, the effect of proactiveness on the relationship between strategic alignment and performance (PROAC × SA → PERF) remains statistically significant and robust across both estimation methods ($\beta = .140$, $T = 2.402$, $p < .05$; $\beta = .233$, $T = 2.132$, $p < .05$), confirming H8. Table III reports the results.

## TABLE III: PLS-SEM MODELS' RESULTS

| Hypotheses | Relationships | Path analysis model β | T-value | PCA analysis model β | T-value | Results |
|---|---|---|---|---|---|---|
| H1 | HAS (QE) → SA | -.149** | 2.405 | -.044** | 1.962 | Supported |
| H2 | FR (QE) → SA | -.173*** | 2.783 | -.072*** | 2.686 | Supported |
| H3 | TU → SA | -.076 | 1.091 | -.032 | 1.014 | Not supported |
| H4 | SA (QE) → PERF | -.124** | 2.515 | -.382** | 2.243 | Supported |
| H5 | RT x HAS → SA | .215*** | 2.961 | .097*** | 3.089 | Supported |
| H6 | RT x TU → SA | .033 | .482 | .031 | 1.012 | Not supported |
| H7 | PROAC x FR → SA | .159 | 1.560 | .084* | 1.773 | Partially supported |
| H8 | PROAC x SA → PERF | .140** | 2.402 | .233** | 2.132 | Supported |
| Controls | EMP → SA | -.036 | .432 | -.001 | .011 | |
| | EMP → PERF | .153*** | 3.074 | .103*** | 2.853 | |
| | REVENUE → SA | .075 | .994 | .038 | .994 | |
| | REVENUE → PERF | -.129* | 1.876 | -.118* | 1.897 | |

Notes: * = $p < .10$, ** = $p < .05$, *** = $p < .01$; Adj. $R^2$ Path model: SA = .367, PERF = .518; Adj. $R^2$ PCA: SA = .343, PERF = .506. The bootstrapping procedure was replicated with 10,000 subsamples, yielding consistent results with those based on 5000 resamples.

**FIG. 2: GRAPHICAL REPRESENTATION OF EO MODERATION EFFECTS**

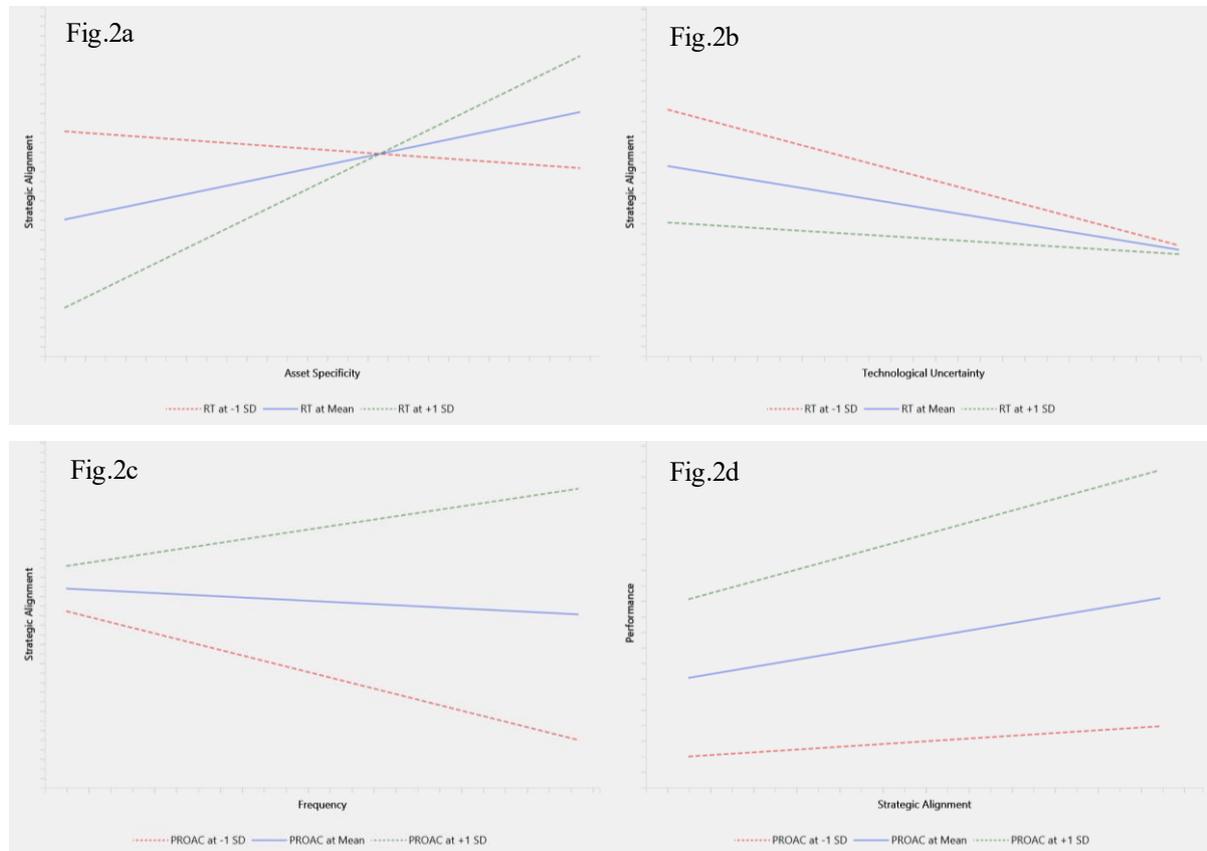

## 7. Study 2

To assess the robustness and generalisability of the findings from Study 1, we conducted a complementary follow-up study (Study 2) adopting a quasi-experimental design based on secondary firm-level data. While Study 1 relies on perceptual measures of governance-enabled performance, Study 2 adopts an alternative operationalisation based on objective, firm-level growth indicators. Specifically, we focus on SaaS adoption for direct e-commerce activities by identifying the degree of strategic alignment associated with Shopify usage across four industries and testing its relationship with long-term firm indicators data from European firms.

### 7.1 Pre-study

The pre-study was administered on the Prolific platform and involved 100 managers currently working in European for-profit companies. To ensure response quality, participants were pre-screened to meet the following criteria: fluent in English; employed in organisations with at least 11 employees; holding a managerial position (from middle management to C-level); and possessing a minimum of an undergraduate degree. Each participant was asked to rate, on a 7-point scale (1 = 'not at all strategic', 7 = 'extremely strategic'), 'How strategically central a direct e-commerce channel would be for the following sector.' The four sectors were selected based on their prevalence in our main dataset and correspond to the Standard Industrial Classification (SIC) codes 56 (Apparel and Accessory Stores), 58 (Eating and Drinking Places), 59 (Miscellaneous Retail), and 73 (Business Services). The aggregated mean scores obtained from this pre-study were used to create an ordinal index of strategic alignment. The index indicates SIC 58 in last place, SIC 73 in third, SIC 59 in second, and SIC 56 in first place. This index reflects the degree to which the e-commerce SaaS adoption can be considered strategically aligned with the core business of firms in each sector. In our quasi-experimental framework, this sector-based alignment variable acts as the independent variable.

### 7.2 Data collection

The dataset for Study 2 was built using secondary information retrieved from Lusha, a professional business intelligence platform that provides verified company-level data (Ballerini *et al.*, 2025; Mathew *et al.*, 2024). We included only firms founded from 2010 onwards, with headquarters located in Europe, and with Lusha's detected adoption of Shopify as their SaaS technology for e-commerce activities. Additionally, we considered only firms for which information on revenues, employee count, and number of funding rounds was available in the database. After applying these criteria, the final sample consisted of 238 firms.

**7.3 Measurements**

Strategic alignment was defined as the degree to which the adopted SaaS – in this case, Shopify, for the direct e-commerce channel – corresponds to and supports the firm's core business activities. In this quasi-experimental study, strategic alignment was operationalised at the industry level using an ordinal scale derived from the Prolific pre-study. This scale reflects the degree of strategic centrality of direct e-commerce for firms operating in each of the four SIC sectors considered. It thus captures an exogenous, sector-based gradient of strategic alignment between e-commerce adoption and firms' typical business logic. The dependent variable – firm performance – was constructed as a composite index that reflects the multidimensional nature of company outcomes, complementing the reflective performance measures in Study 1. Composite performance indices based on firm size, revenue, and funding dynamics are commonly used in entrepreneurship and innovation research as proxies for growth-orientated performance, especially when objective accounting data are unavailable or heterogeneous across firms (Coad *et al.*, 2016; Colombo and Grilli, 2010). The index integrates three dimensions – revenue, employees, and funding rounds – each retrieved from the Lusha database. For comparability, each component was normalised to a common ordinal scale and then mean-centred. The final Performance Index (PERF) was computed as the arithmetic mean of the standardised components, providing a balanced measure of firm performance across economic and structural dimensions. While operationalised differently, this composite index captures growth-orientated firm outcomes that complement the governance performance construct used in Study 1. In addition, several control variables were introduced to account for firm-level heterogeneity. These included the firm's founding year and its country of headquarters – classified into four categories: France; Germany; the United Kingdom; and other European countries.

**7.4 Results**

We first examined the form of the relationship between strategic alignment and firm performance. The curve estimation procedure in SPSS revealed that the quadratic model provided a significantly better fit than the linear one ($R^2 = 0.037$, $F = 4.565$, $p = 0.011$). In contrast, the linear specification was non-significant ($R^2 = 0.001$, $F = 0.261$, $p = 0.610$). The quadratic equation displayed a positive linear term ($b_1 = 0.167$) and a negative squared term ($b_2 = -0.291$), confirming an inverted U-shaped relationship between strategic alignment and performance. Figure 3 illustrates this curvilinear pattern. To verify robustness, we estimated an ordinary least squares (OLS) regression model that included the same independent variable and added country and founding-year controls. The results (summarised in Table IV) stayed consistent with the curvilinear pattern: the squared alignment term remained negative and significant (β = –.148, p = .018). At the same time, multicollinearity diagnostics indicated no concern (all VIF < 1.3). Among the controls, the firm founding year showed a negative and significant effect on performance (β = –0.312, p < 0.001), suggesting that more recently established firms tend to report lower performance scores – an expected outcome, considering their earlier development stage. Country dummies were non-significant, indicating no major cross-national bias. Figure 3 illustrates this curvilinear pattern: performance increases as strategic alignment rises from low to moderate levels, then declines once alignment becomes excessive. Collectively, these findings corroborate the inverted U-shaped association identified in Study 1, thereby supporting H4 again.

**FIGURE 3: CURVILINEAR PATTERN IN STUDY 2**

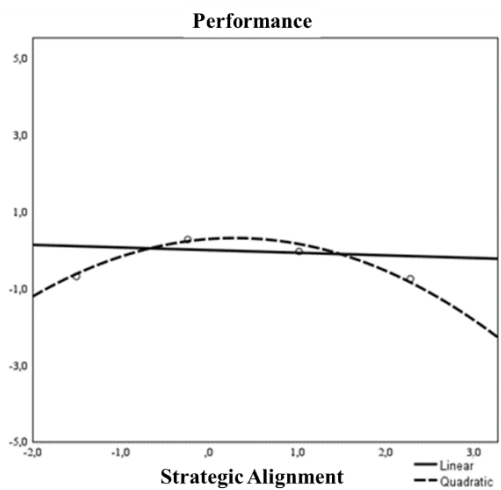

**TABLE IV: OLS RESULTS OF STUDY 2**

| Variables | β | Std. Err. | T value | Sig. | VIF |
|---|---|---|---|---|---|
| (Constant) |  | .371 | -5.010 | <.001 |  |
| Strategic Alignment | -.148 | .082 | -2.373 | .018 | 1.042 |
| FRANCE | -.064 | .519 | -.988 | .324 | 1.130 |
| GERMANY | .073 | .391 | 1.091 | .277 | 1.214 |
| UK | -.121 | .289 | -1.743 | .083 | 1.286 |
| YEAR | -.312 | .041 | -4.931 | <.001 | 1.076 |

*Note: Dependent variable: PERF; Observations 238; Adj $R^2$ =.12*

## 8. Discussion

SaaS adoption has profoundly transformed traditional outsourcing approaches by offering a new model that covers both firms' core and non-core business operations (Cho and Chan, 2015). This model is increasingly adopted in firms in order to achieve business objectives and develop organisational competitive advantages (Attaran and Woods, 2019; Cho and Chan, 2015; Loukis *et al.*, 2019). However, the implementation of SaaS often forces firms to transform their current IT governance and related decision-making process as, based on the model architecture and characteristics, a growing part of the decisions related to IT strategy would be made by instances (i.e. the vendors) outside the firm (Aubert and Rivard, 2020; Benlian and Hess, 2011). This loss of control may pose a potential risk to firms, affecting their core governance-enabled strategic outcomes. Thus, this study explores the subject of SaaS as a hybrid form of governance and its relationship with SMEs' core business performances. Viewed from the TCT perspective, we further examine the relationships and moderating factors (i.e. entrepreneurial and business factors) between SaaS adoption and SMEs' core business performances. This research contributes to ongoing scholarly discussions on the 'hybrid' model of platform governance (Cuypers et al., 2021) and on firms' decision-making alternatives regarding SaaS adoption (Loukis *et al.*, 2019; Schneider and Sunyaev, 2016).

First, our findings (H1 and H2) support the notion that the connection between human asset specificity and frequency, as well as the alignment of the adopted SaaS model with SME core strategy, is explained by a quadratic inverted relationship (inverted U-shaped). In other words, up to a certain point, human asset specificity enables the firm to align SaaS adoption more closely with its core strategy. However, beyond a certain point, the specificity of deploying human assets to manage the SaaS platform makes them less suitable for strategic purposes. As for the frequency, up to a certain point SaaS adoption aligns with SMEs' core strategy. However, suppose it needs to be used frequently in daily tasks. In that case, the study reveals that firms are adopting them for secondary business-related functionalities rather than tasks aligned with the firm's core strategic scope. With this finding, we are enriching the TCT literature by finally responding to the call from Cuypers et al. (2021), which stresses the need to explore digital platforms as a hybrid form of governance.

Second, what emerges from testing and validating H4 is more counterintuitive in relation to the prior literature on the topic. Indeed, even though SaaS is increasingly adopted and leveraged by firms to support their performance (Attaran and Woods, 2019; Loukis et al., 2019), long-term performance seems to drop beyond a certain threshold if SMEs' core strategy depends on SaaS, as supported by Study 1. Although not directly addressing the issue, Benlian and Hess (2011) list strategic and performance risks – such as potential system outages, instability of Internet connectivity, or loss of innovative capacity – that may impact SaaS systems and, thus, firms' performances over time, which can contribute to the overall explanation. Importantly, these findings should be interpreted in terms of governance-enabled long-term performance rather than short-term financial outcomes, highlighting how excessive reliance on SaaS for strategically critical activities may undermine strategic autonomy and adaptive capacity over time.

In addition, a recent study by Alemayehu, Tveteraas, and Kumbhakar (2023) demonstrates that the adoption of enterprise resource planning (ERP – an SaaS system) improves management practices and reduces costs progressively; however, these benefits eventually plateau, as cost minimisation depends on the service quality level. Drawing a parallel with our research, further investigations should delve into the factors (e.g. SaaS quality, flexibility, and vendor responsiveness) that explain this quadratic inverted relationship.

Our Study 2 further corroborates this interpretation using objective firm-level data. This convergence across methods and samples suggests that the relationship between SaaS alignment and long-term governance performance is not only perceptual but also observable in real-world firm outcomes. From a TCT perspective, this turning point reflects the balance between coordination efficiency and adaptive flexibility: when SaaS becomes overly embedded in the firm's strategic core, its standardised and externally governed nature constrains responsiveness, increases dependency on vendors, and erodes the firm's ability to reconfigure its resources autonomously, generating the so-called platform dependency and vendor look-in effects (Cutolo and Kenney, 2021; Zhu and Zhou, 2012). Conversely, when SaaS remains too peripheral, firms fail to capture its potential to enhance efficiency and scalability. Thus, both studies converge in indicating that SaaS adoption contributes most to performance at moderate levels of alignment, where firms can benefit from coordination gains without sacrificing strategic autonomy. In addition, rooted in Braune *et al.*, (2025), Costa *et al.* (2024) and Wu and Pambudi (2024), our results might be nuanced depending on the specificities – such as industry context and digital maturity – of the firms under study, which opens up further avenues for research.

Third, regarding the moderation effect of EO, we validated hypotheses 5, 8, and partially 7. The more firms are willing to take risks in SaaS human assets investments – e.g. hiring, training dedicated staff – the more they will achieve a good alignment between SaaS and SMEs' core strategy directing the firm activities toward high payoff endeavours (see Fig.2a). This EO dimension can therefore be considered a dynamic capability (Dushnitsky and Stroube, 2021; Vial, 2019). Additionally, this result confirms that financial economics theories, which postulate a positive relationship between risk-taking and returns, also hold in entrepreneurship (Putniņš and Sauka, 2020). With regard to H7 and its statistically controversial results, we can clearly see from the interaction plot (Figure 2.c) that for high levels of proactiveness, the relationship between the frequency of SaaS use and its strategic alignment turns markedly positive, while for low levels of proactiveness it becomes negative. This result indicates that frequent SaaS use supports strategic alignment only when entrepreneurs display a proactive posture that fosters experimentation and adaptation of SaaS functionalities to core business processes. Conversely, when proactiveness is low, high usage intensity may merely reinforce dependence on standardised tools, leading to rigidity and misalignment. These findings align with entrepreneurial orientation theory, which posits that proactiveness enables firms to transform technology usage from an operational support activity into a strategic capability (Hervé et al., 2020; Kraus et al., 2023). In this sense, proactiveness can be regarded as a dynamic capability (Dushnitsky and Stroube, 2021; Vial, 2019) that allows SMEs to sense and seize opportunities by leveraging digital frequency into strategic value. As for Hypothesis 8, it highlights the moderating effect of proactivity on SaaS alignment with SME long-term governance performance (see Fig. 2d), aligning with our examination of the literature (Attaran and Woods, 2019; Hervé *et al.*, 2020; Maravić *et al.*, 2022).

Finally, hypotheses 3 and 6, both related to technological uncertainty, were not statistically supported; however, the observed trends aligned with theoretical expectations. In the case of H3, the negative but non-significant association suggests that greater technological uncertainty tends to weaken the alignment between SaaS adoption and SMEs' strategic objectives. This outcome may partly reflect the decision to focus exclusively on technological uncertainty, a deliberate choice made to maintain theoretical coherence with the SaaS context, where uncertainty primarily stems from software evolution, vendor dependency, and limited transparency regarding future updates (Benlian and Hess, 2011; Schneider and Sunyaev, 2016). Nonetheless, this narrow focus may not fully capture how broader environmental or managerial uncertainties interact with technological ones, potentially diluting the statistical effect.

As for H6, the moderation of risk-taking on the technological uncertainty–alignment relationship shows a weakly positive but non-significant coefficient, though the interaction plot reveals a uniformly negative slope across all levels of risk-taking (see Fig. 2b). This pattern implies that increasing technological uncertainty generally reduces strategic alignment, yet highly risk-orientated entrepreneurs seem slightly more capable of containing this decline compared with those exhibiting moderate or low risk-taking. One possible interpretation for this is that, under high technological uncertainty, risk-taking may serve as a compensatory mechanism, allowing firms to absorb disruptions more flexibly or to reinterpret misalignment as exploratory learning. Conversely, when technological uncertainty is low or moderate, high risk-taking appears less beneficial – possibly because entrepreneurs perceive SaaS as a low-risk domain and therefore deploy it in peripheral or experimental functions, weakening its alignment with core strategy. Such findings suggest a context-contingent rather than linear effect of risk-taking, which warrants further exploration through qualitative inquiry.

## 8.1. Theoretical contributions

The theoretical contribution of this paper is manifold. First, the paper joins the discussion of a growing number of studies that highlight digital tools as a suitable model to support SMEs' core business performances (Attaran and Woods, 2019; Beynon *et al.*, 2021; Cho and Chan, 2015; Hervé *et al.*, 2020; Loukis *et al.*, 2019; Yang and Yee, 2022). Second, this study brings originality and novelty to the stand by focusing the analysis on a specific type of digital technology, namely SaaS, and considering its hybrid model of governance, which has remained somewhat neglected (Cuypers *et al.*, 2021). Even more compelling is the initial assumption – that is, considering SaaS's alignment with the SME strategy as a non-linear relationship (Loukis *et al.*, 2019).

Fourth, from a theoretical standpoint, this paper successfully incorporates three dimensions of the TCT perspective (i.e. human asset specificity, frequency of use, and technological uncertainty) into the logic behind the hypothesis formulation that underpins the tested model. Moreover, summoning the EO literature (Kraus *et al.*, 2023; Putniņš and Sauka, 2020), the study brings novelty and development to both the digital transformation and entrepreneurship literature while advancing the understanding of the 'hybrid' governance model of the transaction cost theory (Cuypers *et al.*, 2021). In doing so, we pioneer the bridging between TCT, governance and EO literature. In particular, by applying the EO perspective in the SaaS context, we extend the discussion initiated by Dutta et al. (2024) on the factors influencing SaaS adoption by establishing relationships and identifying moderating factors between SaaS and SMEs' core business performances. Last, but not least, this research complements the discussion initiated by Park *et al.*, (2017) on the effect of internal and external IT governance on distinctive governance-enabled performance.

## 8.2. Practical implications

The adoption of several significant digital technologies has transformed entrepreneurial dynamics, generating multifaceted managerial and policy implications (Lerro *et al.*, 2022). This paper educates practitioners by providing several practical implications for those who wish to adopt the SaaS model. First, the top executives in SMEs' IT governance who wish to support their core business performance through the implementation of SaaS are encouraged to consider the results of this study to define their strategy and support their performance predictions (Cuypers *et al.*, 2021; Loukis *et al.*, 2019; Schneider and Sunyaev, 2016). Indeed, by considering the SaaS hybrid model of governance, our research has produced a comprehensive set of opportunities and challenges to be considered in the decision to adopt SaaS. For instance, a firm selling physical products and offering remote maintenance as a side service to its customers can leverage the SaaS model to a certain extent. Indeed, remote maintenance has become an integral part of the firm's core business, and the frequency of use and the human asset specificity deployed have also increased. As a result, the firm may need to develop an ad hoc solution to reduce transaction costs (i.e. the fee per product monitored). The research has also uncovered the moderating effects of EO. Therefore, before adopting or rejecting SaaS, we strongly encourage managers to perform a preventative evaluation of their level of EO (Fayolle *et al.*, 2010) to modulate their SaaS adoption strategy. Finally, the results provide valuable guidelines for policymakers and governments, particularly those from countries where the adoption of digital technologies is in its early stages (Yang and Yee, 2022). Indeed, they can develop support programmes and training to encourage decisions that lead to SaaS adoption. For instance, in Europe, policymakers can leverage entrepreneurial ecosystems and use them as a catalyst to train and inform SMEs with regard to SaaS.

### 8.3. Research limitations

Although our research makes important contributions, it is not free from limitations which pave the way for future research. First, our study focused on SaaS as a hybrid form of governance within individual SMEs, neglecting how SME–vendor power dynamics and ecosystem orchestration shape governance outcomes (Ballerini *et al.*, 2024). For instance, depending on the vendor policy toward SaaS evolution over time (Benlian and Hess, 2011; Cho and Chan, 2015), more prominent players' requests for evolution may be prioritised and rolled out to all client firms (Gupta *et al.*, 2024), thus impacting SMEs' ability to consolidate their training and funding capacity to adapt to these changes. Therefore, future research can analyse hybrid governance relationship dynamics in an ecosystem. Second, our research did not, per se, consider the lock-in induced by the SaaS switching cost (Cutolo and Kenney, 2021; Zhu and Zhou, 2012). Considering the quadratic inverted relationship (inverted U-shaped) revealed in H1 and H2, this is of interest to understand the consequences for SMEs after the breaking point. Third, despite the adoption of a multimethod approach with two different study settings and corroborated results, there remain structural limitations in the cross-sectional data used in this research.. To capture the decay or reinforcement of SaaS alignment over time, future research could benefit from a longitudinal approach (e.g. time series methodology designs). This would enable us to fully comprehend the sequencing of SaaS adoption and thus identify the critical points of SaaS alignment with SME strategy in various contexts and depending on SMEs' IT governance maturity (Braune *et al.*, 2025; Costa *et al.*, 2024; Wu and Pambudi, 2024). Finally, this study intentionally focuses on specific dimensions of broader constructs due to theoretical coherence and data availability constraints. In particular, uncertainty was operationalised in its technological form – most salient in SaaS governance contexts – while the multidimensional nature of entrepreneurial orientation was reduced to its most empirically relevant facets (risk-taking and proactiveness). Similarly, the performance construct was defined in regard to the long-term to ensure comparability and conceptual clarity; nevertheless, the complementary validation study (Study 2) examined broader performance indicators, including workforce growth and investor funding rounds. While this study deliberately focuses on long-term governance performance to ensure conceptual consistency with IT governance and transaction cost theory, future research could extend this framework by incorporating more specific financial, innovation-related, or resilience-orientated performance outcomes. Future research could expand these constructs to incorporate additional subdimensions and explore their interplay using other methodologies (e.g. case studies, interviews).

## 9. Conclusion

The purpose of this research is to investigate the impact of SaaS adoption (as a 'hybrid' form of governance described in the transaction cost theory, TCT) on SMEs' performance while investigating the pivotal role of SMEs' entrepreneurial orientation (EO) in shaping both SaaS adoption and SME performance outcomes.

Building on a quantitative method, the study reveals that the relationship between two key attributes (namely, asset specificity and frequency) and the alignment of the adopted SaaS model with the firm's core business strategy is explained by a quadratic inverted relationship (inverted U-shaped). In addition, it sheds light on the moderating effect of the risk-taking dimension of the EO construct on asset specificity. Last but not least, the relationship between SaaS alignment with SME strategy and long-term performance is also explained by an inverted U-shaped relationship moderated by EO's proactivity dimension.

Through our results and discussions, we enrich the digital transformation and entrepreneurship literature by comprehending the relationship between SaaS platforms as a hybrid form of governance and SMEs' core business performances, while positioning EO as a moderator. In doing so, we pioneer the bridging between TCT, governance and EO literature and introduce a novel contingency perspective: the effectiveness of governance-enabled strategic alignment in promoting SaaS adoption is shaped by the firm's EO. Last, but not least, from a practicioners perspective, this paper provides guidance to the SMEs' executive and managers who wish to adopt the SaaS model to support their core business performances.

We hope that our research will encourage other scholars to explore this topic as much remains to be investigated to comprehend the whole power dynamics of SaaS as a hybrid form of governance and the role of EO in diverse SME performance metrics over time.

**Acknowledgements**

The authors acknowledge their respective contributions to this research in accordance with the Contributor Roles Taxonomy (CRediT) developed by the National Information Standards Organization (NISO). The specific contributions of each author follow. Jacopo Ballerini: Conceptualization, Methodology, Project administration, Data curation, Formal analysis, Investigation, Software, Visualization, Writing – original draft, Writing – review & editing; Magali Pino: Writing – original draft, Writing – review & editing; Michal Kuděj: Supervision; Alberto Ferraris: Supervision, Funding acquisition. The authors acknowledge the use of ChatGPT-5.2 (OpenAI) and Quillbot (premium version) for copy-editing and language refinement of the manuscript. All suggested edits were reviewed, evaluated, and incorporated manually by the authors. The authors retain full responsibility for the content, interpretations, and accuracy of the final manuscript.

# Appendix A: Discriminant Validity and Robustness Check

The discriminant validity of the measurement model was assessed using the Heterotrait–Monotrait (HTMT) ratio. All HTMT values were below the threshold of 0.90, except for three marginal cases (HAS–FR = 0.915, PERF–FR = 0.915, and RT–PROAC = 0.959). These values are theoretically plausible given the conceptual proximity of constructs such as Risk-Taking and Proactiveness (dimensions of Entrepreneurial Orientation), and the natural linkage between Frequency, Human Asset Specificity, and Performance within SaaS governance contexts. Following Voorhees et al. (2016), such minor exceedances do not compromise discriminant validity, especially when supported by theoretical justification and low multicollinearity (all VIF < 3.0). To further test robustness, an alternative specification was estimated by excluding one indicator each from Risk-Taking and Performance, and incorporating a theoretically validated indicator from De Vita et al. (2010) for Human Asset Specificity. In this reduced model, all HTMT ratios were below 0.90, and structural path results remained consistent with the main model; moreover, H7 gets even better support in respect with the main model ($\beta$ = .189; p = .045). The following tables (A2) report HTMT ratios for both models and path coefficients for the appendix reduced model (A1).

## Table A1. PLS-SEM Path Coefficients – Reduced Model

| Path relationships | β | St. Dev. | T Value | P value |
|---|---|---|---|---|
| HAS (QE) → SA | -.130 | .062 | 2.097 | .036 |
| FR (QE) → SA | -.172 | .063 | 2.730 | .006 |
| TU → SA | -.084 | .067 | 1.257 | .209 |
| SA (QE) → PERF | -.133 | .050 | 2.650 | .008 |
| RT x HAS → SA | .213 | .067 | 3.167 | .002 |
| RT x TU → SA | .041 | .070 | .587 | .557 |
| PROAC x FR → SA | .189 | .094 | 2.006 | .045 |
| PROAC x SA → PERF | .117 | .059 | 1.997 | .046 |
| REVENUE → PERF | -.101 | .066 | 1.518 | .129 |
| REVENUE → SA | .064 | .079 | .815 | .415 |
| EMP → PERF | .155 | .054 | 2.886 | .004 |
| EMP → SA | -.051 | .081 | .630 | .529 |

**TABLE A2: HTMT MODELS COMPARISON**

| Construct Pair | HTMT (Main Model) | HTMT (Reduced model) |
| --- | --- | --- |
| F1 ↔ EMP | 0.204 | 0.198 |
| HAS ↔ EMP | 0.412 | 0.401 |
| HAS ↔ FR | 0.915 | 0.861 |
| PERF ↔ EMP | 0.256 | 0.249 |
| PERF ↔ FR | 0.915 | 0.872 |
| PERF ↔ HAS | 0.884 | 0.853 |
| PROAC ↔ EMP | 0.270 | 0.266 |
| PROAC ↔ FR | 0.751 | 0.742 |
| PROAC ↔ HAS | 0.871 | 0.842 |
| PROAC ↔ PERF | 0.787 | 0.752 |
| REVENUE ↔ EMP | 0.596 | 0.582 |
| REVENUE ↔ FR | 0.036 | 0.031 |
| REVENUE ↔ HAS | 0.303 | 0.298 |
| REVENUE ↔ PERF | 0.117 | 0.108 |
| REVENUE ↔ PROAC | 0.183 | 0.175 |
| RT ↔ EMP | 0.208 | 0.205 |
| RT ↔ FR | 0.730 | 0.715 |
| RT ↔ HAS | 0.811 | 0.789 |
| RT ↔ PERF | 0.758 | 0.731 |
| RT ↔ PROAC | 0.959 | 0.884 |
| RT ↔ REVENUE | 0.139 | 0.132 |
| SA ↔ EMP | 0.284 | 0.279 |
| SA ↔ FR | 0.606 | 0.609 |
| SA ↔ HAS | 0.828 | 0.801 |
| SA ↔ PERF | 0.677 | 0.665 |
| SA ↔ PROAC | 0.536 | 0.520 |
| SA ↔ REVENUE | 0.265 | 0.259 |
| SA ↔ RT | 0.393 | 0.382 |
| TU ↔ EMP | 0.220 | 0.213 |
| TU ↔ FR | 0.756 | 0.744 |
| TU ↔ HAS | 0.745 | 0.731 |
| TU ↔ PERF | 0.892 | 0.869 |
| TU ↔ PROAC | 0.646 | 0.634 |
| TU ↔ REVENUE | 0.150 | 0.143 |
| TU ↔ RT | 0.618 | 0.606 |
| TU ↔ SA | 0.533 | 0.529 |

## Appendix B: Study 1 Items' questions and scales

| Items | Retrieved From: | Item Description | | Scale |
|---|---|---|---|---|
| HAS1 | (De Vita et al., 2010) | I have invested considerably in the training of personnel for the purpose of the SaaS adoption for the business venture I am running. | | 1-7 |
| HAS2 | | Myself and my collaborators have acquired new knowledge in order to adapt to the specific features of the adopted SaaS. | | 1-7 |
| FR1 | (Crook et al. 2013) | We use all our SaaS platforms daily to support business operations. | | 1-7 |
| FR2 | | Our SaaS platforms are all integrated into our organization's daily workflows. | | 1-7 |
| TU1 | (Song and Montoya-Weiss, 2001) | (Reverse) The performance of SaaS providers is easily predictable and monitorable. | | 1-7 |
| TU2 | | (Reverse) The SaaS architecture we use is highly stable and does not require frequent changes. | | 1-7 |
| PROAC1 | (Covin and Wales, 2012) | While running my business, I always try to take the initiative in every situation (e.g., against competitors, in projects when working with others). | | 1-7 |
| PROAC2 | | In the business venture I am running, we excel at identifying opportunities. | | 1-7 |
| PROAC3 | | In the business venture I am running, we initiate actions to which other organizations respond. | | 1-7 |
| RT1 | | The term "risk taker" is considered a positive attribute for people working for the business venture I am running. | | 1-7 |
| RT2 | | My employees and collaborators are encouraged to take calculated risks with new ideas. | | 1-7 |
| RT3 | | The business venture I am running emphasizes both exploration and experimentation for opportunities. | | 1-7 |
| SA1 | (Wu et al., 2015) | Our company prioritizes building strong brand equity to enhance customer trust, loyalty, and perceived value. | Strengthens branding efforts by supporting CRM systems, social media integration, and customer loyalty programs. | 1-5 |
| SA2 | | Our company focuses on minimizing costs and offering competitive pricing as a key differentiator. | Matches cost-driven strategies by enabling automation, efficiency, and scalability for low-cost production or delivery. | 1-5 |
| SA3 | | Our company prioritizes creating unique products or services to establish a distinctive market position. | Supports creating unique offerings by enabling tailored services, analytics, or exclusive features. | 1-5 |
| PERF1 | (Ali and Green 2012) | The SaaS tools we are adopting are well-suited to support my business's long-term growth plans. | | 1-7 |
| PERF2 | | Our adopted SaaS solutions will continue to meet the needs of our business as it scales and expands. | | 1-7 |
| PERF3 | | The SaaS adoption supports the long-term strategic priorities of my business. | | 1-7 |
| PERF4 | | The SaaS tools we use are flexible enough to adapt to future challenges and opportunities in my business. | | 1-7 |
| PERF5 | | Our SaaS adoption will remain cost-effective as my business grows and becomes more complex. | | 1-7 |
| PERF6 | | The SaaS platforms we have adopted will sustain competitive advantages as my business evolves over time. | | 1-7 |